\documentclass[twocolumn,prl,showpacs,amsmath,amssymb]{revtex4}

\usepackage{graphicx}

\def\be{\begin{equation}}
\def\ee{\end{equation}}
\def\bea{\begin{eqnarray}}
\def\eea{\end{eqnarray}}
\def\bma{\begin{mathletters}}
\def\ema{\end{mathletters}}

\def\0{\overline{0}}

\def\q0{\underline{0}}

\def\tr{\mbox{tr}}
\def\one{\leavevmode\hbox{\small1\normalsize\kern-.33em1}}

\begin{document}

\title{Quantum Correlations over Long-distances Using Noisy Quantum Repeaters}

\author{Joonwoo Bae$^{1}$}
\email{bae.joonwoo@gmail.com}
\author{Jeong San Kim$^{2}$}
\email{jkim@qis.ucalgary.ca}

\affiliation{$^{1}$School of Computational Sciences, Korea
Institute for Advanced Study, Seoul 130-012, Korea\\
$^{2}$Institute for Quantum Information Science, University of
Calgary, Calgary, Alberta, Canada}

\date{\today}


\begin{abstract}
Quantum correlations as the resource for quantum communication can
be distributed over long distances by quantum repeaters. In this
Letter, we introduce the notion of a \emph{noisy} quantum
repeater, and examine its role in quantum communication. Quantum
correlations shared through noisy quantum repeaters are then
characterized and their secrecy properties are studied.
Remarkably, noisy quantum repeaters naturally introduce
\emph{private states} in the key distillation scenario, and
consequently key distillation protocols are demonstrated to be
more tolerant.
\end{abstract}


\maketitle

Quantum Key Distribution (QKD) protocols such as the
Bennett-Brassard $1984$ (BB84) \cite{bb84} have been implemented
in laboratories \cite{qkd1,review}, and become one of the most
important and promising applications of Quantum Information
Theory. QKD is now no longer an experiment but an emerging market
\cite{market}. Further investigations on QKD protocols will
improve their practical performance under realistic constraints
\cite{rs}. To date, QKD reaches about $100$ $km$ in distance with
photon sources through optical fibers, which however does not yet
meet the distance standard of present-day communication.

The communication distance is somehow limited as any physical
resource carrying quantum states suffer unwanted interactions with
environment such as decoherence and losses during transmission. In
this sense, it is natural to build a bridge for quantum
correlations, for instance quantum relay or quantum repeater, to
overcome the distance limit. Quantum repeaters are in fact known
to efficiently extend the communication distance \cite{repeater,
gisin np}, but unfortunately not feasible within current
technology since the so-called \emph{quantum memory}, that stores
quantum states for a while, is experimentally challenging.
Nevertheless, there have been remarkable experimental results that
envisage a feasible quantum memory in the near future
\cite{repeaterex}.

This work is therefore motivated by two perspectives. First, a
quantum memory in the near future, as being in an earlier phase of
development, would have a storage-time long enough to distribute
quantum correlations over distances, but not sufficiently long to
apply entanglement distillation. The next arises from the fact
that a practical quantum repeater, being contacts to the quantum
channels, would be susceptible to its surroundings. To be
specific, as quantum repeaters are connected to one another by
possibly noisy quantum channels, errors caused by the noisy
channels will be ported to the quantum state of a quantum memory
of the repeater, i.e. quantum repeaters become noisy. The question
we address then is in what way do \emph{noisy quantum repeaters}
feature in QKD scenarios. It is actually not straightforward to
conclude that the secret key rate decreases, since noise effects
do not always degrade protocols \cite{bgr}. Indeed, we will show
that noisy repeaters degrade the power of eavesdropper, called
Eve. In this Letter, we characterize quantum correlations
distributed through noisy quantum repeaters, and then study
distillation of secret key and entanglement. The distribution
scenario is described in the entanglement-based scheme.

\begin{figure}
  \includegraphics[width=6.5cm]{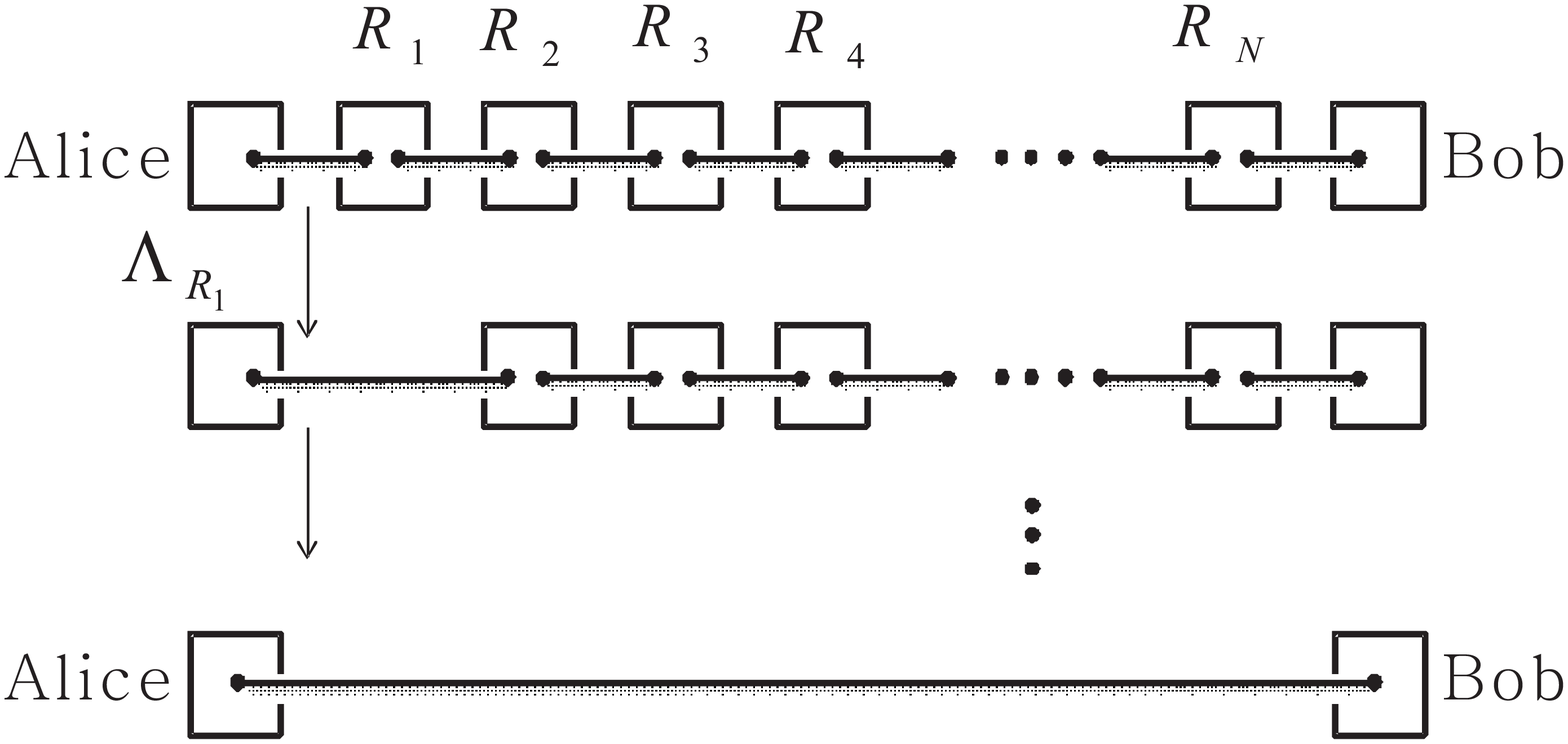}\\
  \caption{Successive applications of $\Lambda_{R_{j}}$ for $j=1,\cdots,N$
  at individual quantum repeaters allow two honest parties to share entangled states.
  }\label{qr}
\end{figure}

We first briefly review entanglement distribution through a single
quantum repeater, denoted by $R$, assuming that all quantum
channels are perfect but only limited in distance. The
distribution scenario follows the standard scheme in Ref.
\cite{repeater}, as follows. Alice first generates the maximally
entangled state $|\phi_{1} \rangle$ where $|\phi_{1} \rangle =
(|00\rangle + |11\rangle)/\sqrt{2}$, keeps the first qubit, and
send the other one to the repeater. Bob does the same, and the
repeater then has two qubits in store. The entanglement swapping
(ES) protocol, denoted by $\Lambda_{R}$, is applied to the two
qubits in store, and afterwards two honest parties share the state
$|\phi_{1}\rangle_{AB}$. Here, the protocol $\Lambda_{R}$ is
composed of Bell-basis measurement on the two qubits followed by
the public announcement of the measurement outcome, in order that
two honest parties apply local operations to rotate the shared
state into $|\phi_{1}\rangle$. A quantum repeater being a quantum
device whose physical state is described by a density operator,
say $\eta_{R}$, as follows, \bea \Lambda_{R}
(|\phi_{1}\rangle_{AR} \otimes |\phi_{1} \rangle_{RB}) =
|\phi_{1}\rangle_{AB}\langle \phi_{1}| \otimes \eta_{R}.
\label{eswap}\eea It has been presumably of no interest to find
out in which quantum state a quantum repeater remains. This is
because of mainly two reasons: as it is seen in (\ref{eswap}), a
repeater is factorized out from two honest parties by the
Bell-basis measurement of the ES protocol, meaning that no secret
correlations between the repeater and two honest parties would be
exploited \cite{CLL}. In addition, it is often supposed that a
repeater stays in a constant state $\eta_{R}$ all the time not
being affected by any change of its surroundings.


We now turn to the realistic constraints to quantum channels and a
quantum repeater. First, quantum channels are noisy in general,
and therefore each of Alice and Bob shares mixed states with the
repeater. Here we restrict to cases where the mixed state is
Bell-diagonal in a single-copy level. If it is not the case, two
honest parties can apply local filtering operations such that
Bell-diagonal states are shared \cite{filter}. The filtering
operation in fact increases the amount of entanglement, in terms
of entanglement of formation \cite{con}, of shared states with
some probability. Possible errors that may happen must be then one
of three kinds, phase-shift, bit-flip, or both. For each case, the
shared state is one of Bell states, where $|\phi_{2}\rangle =
(\openone\otimes Z)|\phi_{1}\rangle$, $|\phi_{3}\rangle =
(\openone\otimes X)|\phi_{1}\rangle$ and $|\phi_{4}\rangle =
(\openone\otimes iY)|\phi_{1}\rangle$ with Pauli matrices $X$,
$Y$, and $Z$.

Next, a practical quantum repeater, as a quantum device, would
react susceptibly to a change of its surroundings. In particular,
a quantum memory in the repeater interacts with two qubits sent by
two honest parties, and thus two qubits stored become the
effective environment of the repeater. Then, suppose that the
two-qubit state encoded by two honest parties are sent through and
perturbed in noisy channels. The isomorphism between quantum
channels and quantum states tells us that all the properties of
the noisy channels can be found in the quantum state that have
arrived at the repeater \cite{iso}. This means that, an error
caused by noisy channels corresponds to a change in the repeater's
environment. In terms of noise parameters, when
$|\phi_{i}\rangle_{AR}$ and $|\phi_{j} \rangle_{RB}$ are shared
for some $i$ and $j$ depending on channel properties, a repeater
would be perturbed according to the noise $i$ and $j$. This is
what we mean that a practical quantum repeater is noisy.

We need to clarify here that it is two honest parties who prepare
and put a repeater in the middle. This means that they already
know the properties of the repeater, how it reacts to each of
phase-, bit-, and both errors, and is perturbed. As well, the ES
protocol in a repeater cannot be designed to work by recognizing
errors instance \emph{per} instance i.e. which pair of Bell-states
is kept, but always assumes the ideal case that $|\phi_{1}
\rangle_{AR} \otimes |\phi_{1} \rangle_{RB}$ is shared. In
addition, we only suppose the minimal responsibilities to a
repeater, performing the ES protocol, and do not consider cases
where repeaters collaborate with two honest parties or an
eavesdropper.

After the ES protocol on $|\phi_{i}\rangle_{AR} \otimes
|\phi_{j}\rangle_{RB}$ in the repeater, two honest parties share
one of Bell states, denoted by $|\phi_{k}\rangle$ where $k$ is
completely determined by the error kinds $i$ and $j$. The noisy
quantum repeater can also be captured by the single index, $k$, as
follows, \bea \Lambda_{R}(|\phi_{i}\rangle_{AR} \otimes
|\phi_{j}\rangle_{RB} ) = | \phi_{k}\rangle_{AB}\langle
\phi_{k}|\otimes \eta_{R}(k). \label{re} \eea The repeater's
quantum state $\eta_{R}(k)$ are known to two honest parties since
they have prepared a repeater. Likewise, in a general case when
$N$ noisy quantum repeaters are located, Bell-diagonal states will
also end up between two honest parties, and the shared state would
be finally of the following form, \bea \rho_{ABR} &=&
\bigotimes_{n=1}^{N} \Lambda(|\phi_{1}\rangle_{R_{n-1}R_{n}}
\otimes |\phi_{1}\rangle_{R_{n}R_{n+1}})
\nonumber\\
&=& \beta_{1} |\phi_{1}\rangle_{AB}\langle\phi_{1}|\otimes\eta_{1}
 + \beta_{2} |\phi_{2}\rangle_{AB}\langle\phi_{2}|\otimes\eta_{2} \nonumber \\
 & + & \beta_{3} | \phi_{3}\rangle_{AB}\langle\phi_{3}|\otimes\eta_{3}
 + \beta_{4}|\phi_{4}\rangle_{AB}\langle\phi_{4}|\otimes\eta_{4}
 \label{state} \eea where $\eta_{j}$ is the normalized quantum
 state of $N$ repeaters when two honest parties share
$|\phi_{j}\rangle$. Interestingly, this is exactly the state that
has been considered in the context of distillation of private
states in Ref. \cite{horo}, except only that the states $\eta_{j}$
do not belong to Alice and Bob but an independent party,
repeaters.

We have characterized quantum states shared through noisy
repeaters and noisy channels, exploiting the independent party,
repeaters, in the shared state. Before starting secrecy analysis,
we should first introduce the important assumption on shared
states, that quantum states shared by two honest parties are made
invariant under any permutation of pairs i.e. symmetric states.
This can be done by random permutations. The quantum de Finetti
theorem then states that the most general $N$-symmetric state,
i.e. invariant under any permutations, $\rho^{(N)}$ converges
efficiently to the identically and independently distributed
($i.i.d.$) one, $\rho_{ABR}^{\otimes N}$, as $N$ becomes a very
large number \cite{renner}. This implies that in the context of
QKD, one does not have to go through the most general case of
$N$-symmetric states $\rho_{ABR}^{(N)}$ but it suffices to
consider the $i.i.d.$ one, the so-called collective attacks,
$\rho_{ABR}^{\otimes N}$ \cite{renner}. The bound for the general
security can be obtained by analyzing collective attacks. In what
follows, we assume that the number of copies, $N$, becomes a very
large number to ensure applicability of the quantum de Finetti
theorem.

We now translate the scenario introduced in Ref. \cite{horo} of
distilling private states to that of distilling secret key with
noisy quantum repeaters. Since repeaters are supposed to comprise
an independent party not belonging to two honest parties nor the
eavesdropper, errors caused by those repeaters have nothing to do
with eavesdropping strategies and therefore do not have to be
necessarily corrected to distill secret key. This also means that
the quantum state representing the general security does not
necessarily correspond to $|\phi_{1}\rangle_{AB}$. As it was
pointed out in \cite{horo}, by introducing a party independent to
two honest parties and Eve, Eve's purification power can be
degraded. All this can be encapsulated by the so-called private
states, corresponding to the general security, as follows, \bea
\gamma_{ABR} = U |\phi_{1}\rangle_{AB} \langle \phi_{1}|\otimes
\rho_{R} U^{\dagger}, \label{gamma}\eea where $\rho_{R}$ is a
repeater state and $U = \sum_{i,j}|ij\rangle_{AB}\langle ij|
\otimes V_{R}^{i,j}$ is a unitary operation called $twisting$.
Note that a twisting $U$ provides the equivalence class of private
states. The only difference is that the local assistance by
$A^{'}$ and $B^{'}$ in Ref. \cite{horo} is replaced with noisy
quantum repeaters.

In particular, we take a unitary operation, the untwisting one
$U^{\dagger}$ in Ref. \cite{other} that works for a state of the
form in (\ref{state}) as follows, \bea \rho_{ABR} = U^{\dagger}
\sigma_{ABR} U\label{sig}\eea where $\sigma_{AB}(=
\tr_{R}[\sigma_{ABR}]) = \sum_{j} \lambda_{j} |\phi_{j}\rangle
\langle \phi_{j}|$ is again Bell-diagonal, with \bea \lambda_{1,2}
& = & \frac{1}{2}(\|\beta_{1}\eta_{1} + \beta_{2}\eta_{2} \|
\pm \|\beta_{1}\eta_{1} - \beta_{2}\eta_{2}\|) \nonumber \\
\lambda_{3,4} & = & \frac{1}{2}(\|\beta_{3}\eta_{3} +
\eta_{4}\beta_{4}\| \pm \|\beta_{3}\eta_{3} -\beta_{4}
\eta_{4}\|), \label{lam} \eea where the trace-norm of an operator
$A$ has been denoted by $\|A \|$ \cite{other}. This untwisting
operation shows the phase errors that are necessary to be
corrected for the $\rho_{ABR}$ to be a private state. Note that
repeaters are classically correlated with two honest parties in
the state $\sigma_{ABR}$, which means that measurement outcomes of
two honest parties are independent to quantum states of repeaters.
The relations above (\ref{sig}) and (\ref{lam}) reveal not only
the reason why not all errors existing in $\rho_{ABR}$ have to be
corrected but also how much errors are necessarily to be
corrected. First, the state that we are aiming to distill is a
private state, for which as it can be seen in (\ref{gamma}) the
state $\sigma_{AB}$ in (\ref{sig}) should be transformed to the
state $|\phi_{1}\rangle$ by correcting all errors there, though
$\rho_{AB} (= \tr_{R}[\rho_{ABR}])$ may still possess some errors.
Next, $\sigma_{AB}$ has less errors than $\rho_{AB}$, which can be
seen by comparing $\beta_{j}$ of $\rho_{AB}$ in (\ref{state}) to
$\lambda_{j}$ of $\sigma_{AB}$ in (\ref{lam}) for $j=1,2,3,4$. To
be precise, it holds that $\lambda_{j} \leq \beta_{j}$ for
$j=2,3,4$ \cite{formula, pri}. Therefore, correcting less errors
that exist in $\sigma_{AB}$, two honest parties will share secret
key related to a private state. The question followed is whether
key distillation techniques such as advantage distillation and the
standard one-way distillation protocol commute with a twisting
operation $U$. In fact, the distillation techniques commute with
twisting operations \cite{formula, pri} and therefore, no more
additional step is required in the key distillation scenario. The
only difference in the classical step is that less errors are
estimated to be corrected.

We now analyze secrecy properties of quantum states in
(\ref{state}) shared through noisy repeaters. Measuring
$\rho_{ABR}^{\otimes N}$ in the computational basis, two honest
parties are with measurement outcomes of probability distribution,
\bea p_{AB}(i,j) = \langle i_{A}j_{B}|\tr_{R}\rho_{ABR}
|i_{A}j_{B} \rangle, \label{prob} \eea to which key distillation
techniques are then applied. Note here the precondition for key
distillation, that quantum state from which secrecy is to be
distilled must be entangled \cite{CLL}. In this case,
$\sigma_{AB}$ in (\ref{sig}) is the state where secrecy is to be
extracted, and once the state $\rho_{ABR}$ is identified through
state tomography, from the relation (\ref{lam}) it can be easily
checked if the state $\sigma_{AB}$ is entangled. The state
$\sigma_{AB}$ is entangled if and only if the follows holds
\cite{con}, \bea \| \beta_{1}\eta_{1} - \beta_{2}\eta_{2}\| > \|
\beta_{3}\eta_{3} + \beta_{4}\eta_{4} \|.\label{ent}\eea It is
also worth mentioning a sufficient condition: if the state
$\rho_{AB}$ is entangled, i.e. $\beta_{1}-\beta_{2} > \beta_{3} +
\beta_{4}$, so is $\sigma_{AB}$. This is clear from the
inequality, $\beta_{1}-\beta_{2} \leq \| \beta_{1} \eta_{1} -
\beta_{2} \eta_{2} \|$, that holds true for all $\beta_{1,2}$ and
$\eta_{1,2}$.

We now consider the standard one-way communication of error
correction and privacy amplification to the measurement outcomes
of the probability distribution (\ref{prob}). A lower bound to the
one-way secret key rate has been derived in Refs. \cite{dw},
$K_{\rightarrow} \geq I_{AB} - I_{AE}$, where the mutual
information is denoted by $I$. It is then straightforward to
compute the lower bound to the key rate for measurement outcomes
in (\ref{prob}), as follows \bea K_{\rightarrow} & \geq & 1- h(x)
-\sum_{i=1,3} \|\beta_{i}\eta_{i} + \beta_{i+1}\eta_{i+1} \|
h(y_{i}) \label{one} \eea where $h(\cdot)$ is the binary entropy,
$x=\|\beta_{1}\eta_{1} + \beta_{2}\eta_{2}\|$ and $ y_{i} = (1 +
\|\beta_{i}\eta_{i} - \beta_{i+1}\eta_{i+1} \| / \| \beta_{i}
\eta_{i} + \beta_{i+1} \eta_{i+1}\|)/2$. One can see from the
security condition (\ref{one}) how the quantum states of quantum
repeaters are relevant. In the following example, we consider the
BB84 protocol with noisy quantum repeaters, and show that noisy
quantum repeaters make the protocol more tolerant.

\begin{figure}
  \includegraphics[width=5cm]{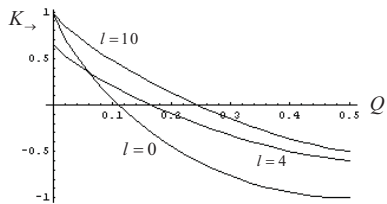}\\
  \caption{The lower bound to the secret key rate is shown for different
  size of repeater states: $l=0$ (without repeaters), $l=4$, and $l=10$. }\label{onekey}
\end{figure}

\emph{\textbf{Example.}}\textbf{(BB84)} In the entanglement-based
scheme, the shared quantum state in the BB84 protocol is
identified in a single-copy level as \bea \rho_{AB} & = &
(1-Q)^{2}|\phi_{1}\rangle\langle\phi_{1}| + Q(1-Q)
|\phi_{2}\rangle \langle\phi_{2}| \nonumber \\ & & + Q(1-Q)
|\phi_{3}\rangle \langle\phi_{3}| + Q^{2} |\phi_{4}\rangle \langle
\phi_{4}| \label{bb84} \eea where $Q$ is the
quantum-bit-error-rate(QBER) \cite{crk}. As a toy model of noisy
quantum repeaters, we here take the shield states considered in
Ref. \cite{horo} as their quantum states, $\eta_{1,3,4} =
(\rho_{s}+\rho_{a})^{\otimes l}/2^{l}$ and $\sigma_{2}  =
\rho_{s}^{\otimes l}$, where $\rho_{a(s)}$ is the normalized
$d$-dimensional projection operator onto asymmetric(symmetric)
space. These noisy quantum repeaters, where only $\eta_{2}$ is
different from the others, can be interpreted as being sensitive
to phase errors. Then, the shared state in a single-copy level can
be written as follows, \bea \rho_{ABR} & = &
(1-Q)^{2}|\phi_{1}\rangle\langle\phi_{1}|\otimes\eta_{1} + (1-Q)Q
|\phi_{2}\rangle \langle\phi_{2}|\otimes\eta_{2} \nonumber
\\ & + & (1-Q)Q |\phi_{3}\rangle \langle\phi_{3}| \otimes\eta_{3}
+ Q^{2} |\phi_{4}\rangle \langle \phi_{4}|\otimes\eta_{4}.
\label{exstate} \eea As we have shown, it suffices to consider
errors existing in the untwisted state $\sigma_{AB}$ in
(\ref{sig}) so that private states in (\ref{gamma}) are to be
distilled. The lower bound to the secret key rate can be computed
from (\ref{one}), and is depicted in Fig. \ref{onekey}. When $l$
becomes a very large number, the QBER converges to $Q = 24.5\%$,
much higher than the known bound $11.0\%$ in Ref. \cite{sp}.

The key distillation technique applying the two-way (AD) followed
by the one-way key distillation tolerates higher values of QBER.
This has been completely analyzed in Ref. \cite{formula}, which
shows that if shared states satisfy \bea \| \beta_{1}\eta_{1} -
\beta_{2} \eta_{2} \|^{2} > \| \beta_{1} \eta_{1} + \beta_{2}
\eta_{2} \| \|\beta_{3} \eta_{3} + \beta_{4}\eta_{4}\|,
\label{twoway}\eea then secret key can be distilled. By this,
quantum states in a wider range are shown to be distilled to
secret key. In particular, if $\eta_{1}$ is orthogonal to
$\eta_{2}$, i.e. $\tr[\eta_{1}\eta_{2}] = 0$, it holds that $\|
\beta_{1}\eta_{1} + \beta_{2}\eta_{2} \| = \| \beta_{1}\eta_{1} -
\beta_{2}\eta_{2}\|$, remarkably meaning that the security
condition in (\ref{twoway}) coincides to the precondition for key
distillation (\ref{ent}). Therefore, for this particular case, all
secret correlations derived from entangled states $\sigma_{AB}$
can be converted to secrecy. To our knowledge, this is the first
case that entanglement itself implies secrecy, although its
general connection remains open.

Finally, we would like to mention that noisy quantum repeaters do
not really play a role in distilling entanglement. This is
because, differently to key distillation where secret key rate
matters, entanglement distillation is concerned with the singlet
fidelity $F = \langle \phi_{1} | \tr_{R}[\rho_{ABR}] |
\phi_{1}\rangle$, where quantum repeaters are traced out and
therefore not considered. However, it would be interesting still
to characterize errors that pertain to the distillation rate of
entanglement.

To conclude, we have characterized quantum correlations
distributed through noisy quantum repeaters, and shown that the
scenario corresponds to the distillation of private states. This
helps to close the gap between two extreme sides of QKD, one
theoretical and the other practical. Remarkably, QKD protocols are
shown to tolerate higher values of QBER by noisy quantum
repeaters. Here the lesson to practical QKD over long distances,
usually having higher values of QBER, is that noise effects
independent to Eve would make more tolerant protocols, for
instance, where noisy ES protocols are included. In addition,
although the scheme of quantum repeater introduced in Ref.
\cite{repeater} is considered, our results can be easily
generalized to other variant schemes that are basically equivalent
to that in Ref. \cite{repeater}.


We would like to thank Yuan Liang Lim for helpful discussions and
comments. This work is supported by the IT R\&D program of
MIC/IITA [2005-Y-001-04, Development of next generation security
technology] and Alberta's informatics Circle of Research
Excellence (iCORE).

\end{document}